\documentstyle[11pt,paspconf,epsf]{article}

\begin{document}

\title{The Shape of the Milky Way's Dark Halo}
\author{R.P. Olling and M.R. Merrifield}
\affil{Department of Physics and Astronomy, University of Southampton,
Southampton SO17 1BJ, England}

\begin{abstract}

The distribution of shapes of galaxies' dark halos provides a basic
test for models of galaxy formation.  To-date, few dark halo shapes
have been measured, and the results of different methods appear
contradictory.  Here, we add to the sample of measured shapes by
calculating the flattening of the Milky Way's dark halo based on the
manner in which the gas layer in the Galaxy flares with radius.  We
also test the validity of this technique -- which has already been
applied to several other galaxies -- by comparing the inferred halo
flattening to that obtained from a stellar-kinematic analysis, which
can only be applied to the Milky Way.  Both methods return consistent
values for the shape of the Milky Way's halo, with a
shortest-to-longest axis ratio for the dark matter distribution of $q
= 0.75 \pm 0.25$.  However, this consistency is only achieved if we
adopt a value of $R_0 = 7 \pm 1\,{\rm kpc}$ for the Sun's distance to
the Galactic center.  Although this value is smaller than the
IAU-sanctioned $R_0 = 8.5\,{\rm kpc}$, it is quite consistent with
current observations.  Whatever value of $R_0$ is adopted, neither
method returns halo parameters consistent with a disk-like mass
distribution, for which $q \sim 0.2$.  This finding rules out cold
molecular gas and decaying massive neutrinos as dark matter
candidates.

\end{abstract}

\section{Introduction}

Cosmological simulations have now advanced to a point where the
formation of individual galaxy halos can be followed (e.g.\ Navarro,
Frenk \& White 1996).  Unfortunately, the processes of star formation
are not yet well enough understood for us to be able to translate
these simulations into what we might expect for the luminous
properties of galaxies.  Thus, if we are to test the cosmological
models, we must determine quantitative properties of dark halos
observationally, and compare these values to the predictions of the
models.  One important discriminant between models is provided by the
shapes of the dark matter halos, which are usually quantified by the
shortest-to-longest axis ratio of the dark matter distribution, $q$.
Cosmological models which are dominated by hot dark matter produce
rather round galaxy halos, with $q \sim 0.8$ (Peebles 1993).
Conversely, in models dominate by cold dark matter (CDM), the halos
that form are significantly flattened: Dubinski (1994) found that the
distribution of $q$ for the ensemble of galaxy halos formed in CDM
simulations could be approximated by a Gaussian with a mean of 0.5 and
a dispersion of 0.15.  Some dark matter candidates such as cold
molecular hydrogen (Pfenniger, Combes \& Martinet 1994) or decaying
massive neutrinos (Sciama 1990) predict that the dark material should
form a flattened disk, resulting in even lower values of $q \sim 0.2$.
Clearly, the true distribution of $q$ for the dark halos around
galaxies offers an important test for cosmological theories.

\begin{figure}
\plotfiddle{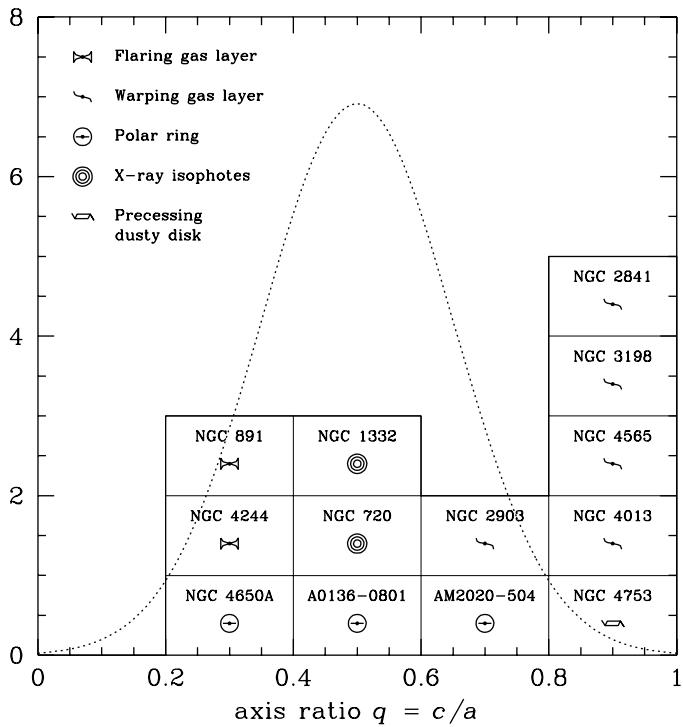}{3.75in}{0}{50}{50}{-150}{-80}
\caption{Histogram of the measured distribution of $q$,
shortest-to-longest axis ratio for galaxy dark matter halos.  The name
of the galaxy and the method employed to measure $q$ are marked in
each block of the histogram.  The dotted line shows the distribution
of $q$ predicted by a cosmological simulation (Dubinski 1994).}
\end{figure}

Unfortunately, measuring a dark halo's shape is a difficult process.
The rotation curve (circular speed as a function of radius) of a
spiral galaxy will tell us about the radial distribution of dark
matter in the system, but it gives few clues as to the mass
distribution's degree of flattening.  Measurements of $q$ therefore
depend on subtler dynamical effects, such as: the manner in which a
galaxy's gas layer warps with radius (e.g.\ Hofner \& Sparke 1994);
the degree to which its gas layer flares with radius (e.g. Olling
1996); the kinematics of the polar rings found around a few galaxies
(e.g.\ Sackett et al.\ 1994); the distribution taken up by the X-ray
emitting gas around elliptical galaxies (e.g. Buote \& Canizares
1996); or, in the case of NGC~4753, detailed modeling of a precessing
dusty disk (Steiman-Cameron, Kormendy \& Durisen 1992).  The results
of such analyses are summarized in Fig.~1, together with the
distribution of $q$ as predicted by a CDM simulation.  As is apparent
from this figure, the difficulty of the requisite analysis means that
there are regrettably few systems for which $q$ has been determined.
Even more disturbing is the fact that different techniques seem to
yield systematically different answers.  The analysis of warped disks,
for example, seems to imply that halos are systematically rounder than
is inferred from the other techniques, while the flaring gas layer
method comes up with systematically flatter halos.

In this paper, we derive the flattening of the dark halo in our own
galaxy, the Milky Way, using the flaring of its gas layer.  This
analysis, presented in Section~2, allows us to add a further datum to
the embarrassingly-sparse histogram in Fig.~1.  Perhaps more
importantly, however, our position within the Milky Way means that we
can also measure $q$ using stellar kinematics (see Section~3) -- a
technique which is unavailable for external galaxies.  Thus, we can
test the methods used to measure $q$ by checking for consistency
between the different approaches; the implications of this comparison
are discussed in Section~4.

\section{The Shape of the Milky Way's Halo from its Flaring Gas Layer}

The manner in which the gas layer in a disk galaxy thickens with
radius provides a conceptually-simple method for measuring the shape
of its dark halo (Olling 1996, and references therein).  The thickness
of the gas layer is dictated by the hydrostatic balance between the
random motions of the gas and the pull of gravity toward the plane of
the galaxy.  Thus, the thickness of the gas layer measures the
gravitational field close to the galaxy's plane, and hence the amount
of dark matter in this region.  For a given total mass of dark matter
(as derived from the galaxy's rotation curve), we can thus derive the
flattening of the halo.  If, for example, the gas is found to be
confined in a very thin layer, then we know that it is being held
there by the gravitational pull of a large amount of dark matter close
to the plane of the galaxy.  The presence of large amounts of dark
matter close to the plane implies that its distribution must be highly
flattened, and hence that $q$ must be small.

In order to apply this method, we need to measure a galaxy's rotation
curve, $\Theta(R)$, and how its gas layer thickness varies with
radius, $h_{\rm gas}(R)$.  For the Milky Way, these two quantities are
intimately entwined.  To convert the observed angular extent of the
gas layer at some point in the Galaxy into a physical thickness, we
need to know the distance to that point.  In general, the only way to
calculate that distance is to adopt a particular rotation curve for
the Galaxy, which will predict how the line-of-sight velocity of gas
in a given direction should vary with the material's distance.  Thus,
the line-of-sight velocity of the gas (as measured from the Doppler
shift in its $21\,{\rm cm}$ emission) can be translated directly into
the requisite distance.  In fact, it is possible to disentangle
$\Theta(R)$ and $h_{\rm gas}(R)$ in the Milky Way by solving
simultaneously for both quantities, providing one of the few direct
methods for calculating the rotation curve of the outer Galaxy
(Merrifield 1992).

The only serious impediment to this analysis is that the derived
functions depend significantly on the values that we adopt for our
distance from the Galactic center, $R_0$, and the circular speed of
the Galaxy at the solar radius, $\Theta_0$.  Observationally, there
are modest uncertainties in both these quantities: $R_0 = 7.8 \pm
0.7\,{\rm kpc}$ (Reid 1993), and $\Theta_0 = 200 \pm 20\,{\rm
km}\,{\rm s}^{-1}$ (Sackett 1997).  However, they are not
independently variable, since local stellar kinematic measurements of
the Oort constants imply that $\Theta_0/R_0 = 26.4 \pm 1.9\,{\rm
km}\,{\rm s}^{-1}\,{\rm kpc}^{-1}$ (Kerr \& Lynden-Bell 1986).
Nevertheless, these uncertainties translate into quite large
variations in the inferred rotation curve and gas layer thickness.

\begin{figure}
\plottwo{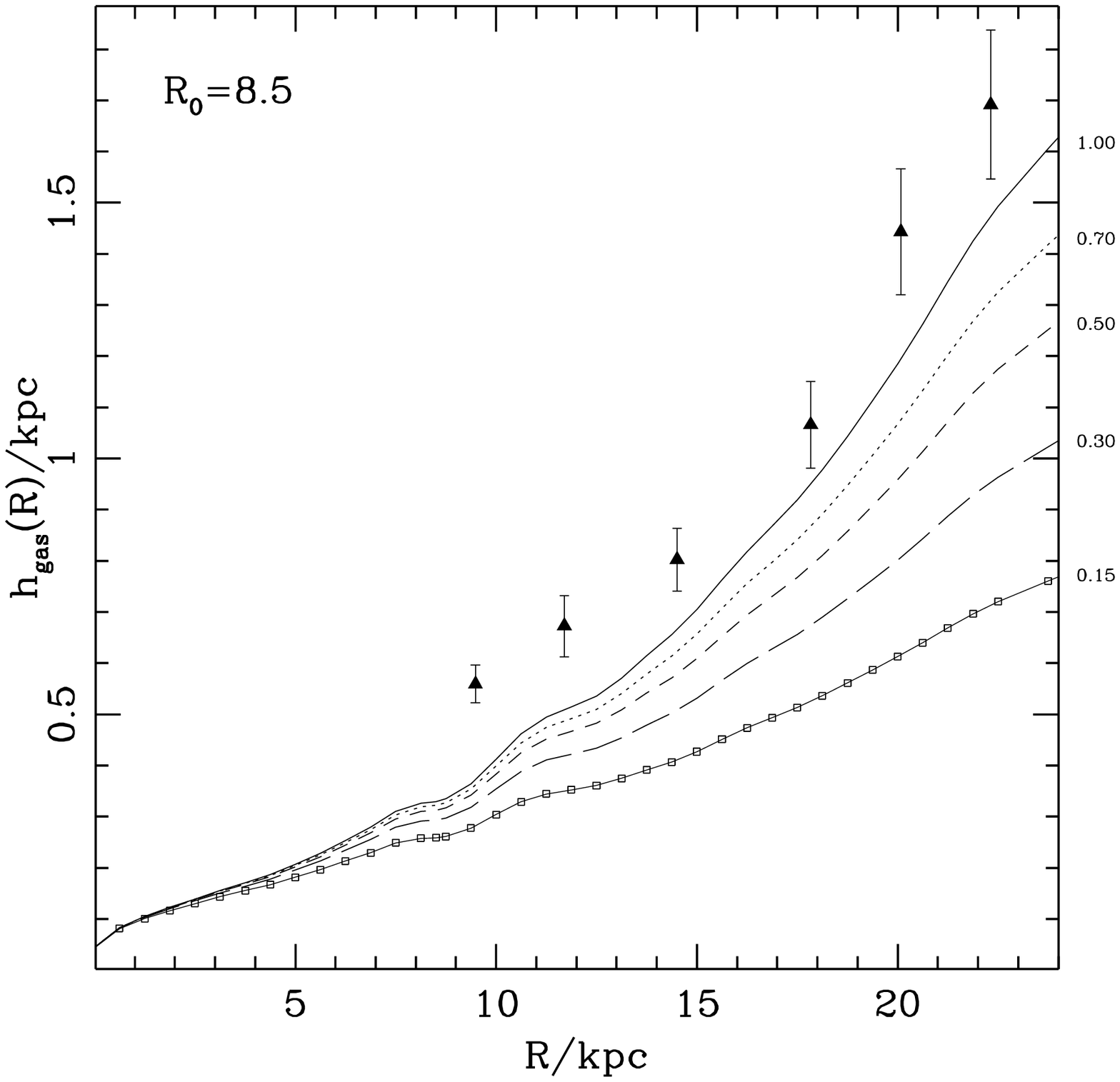}{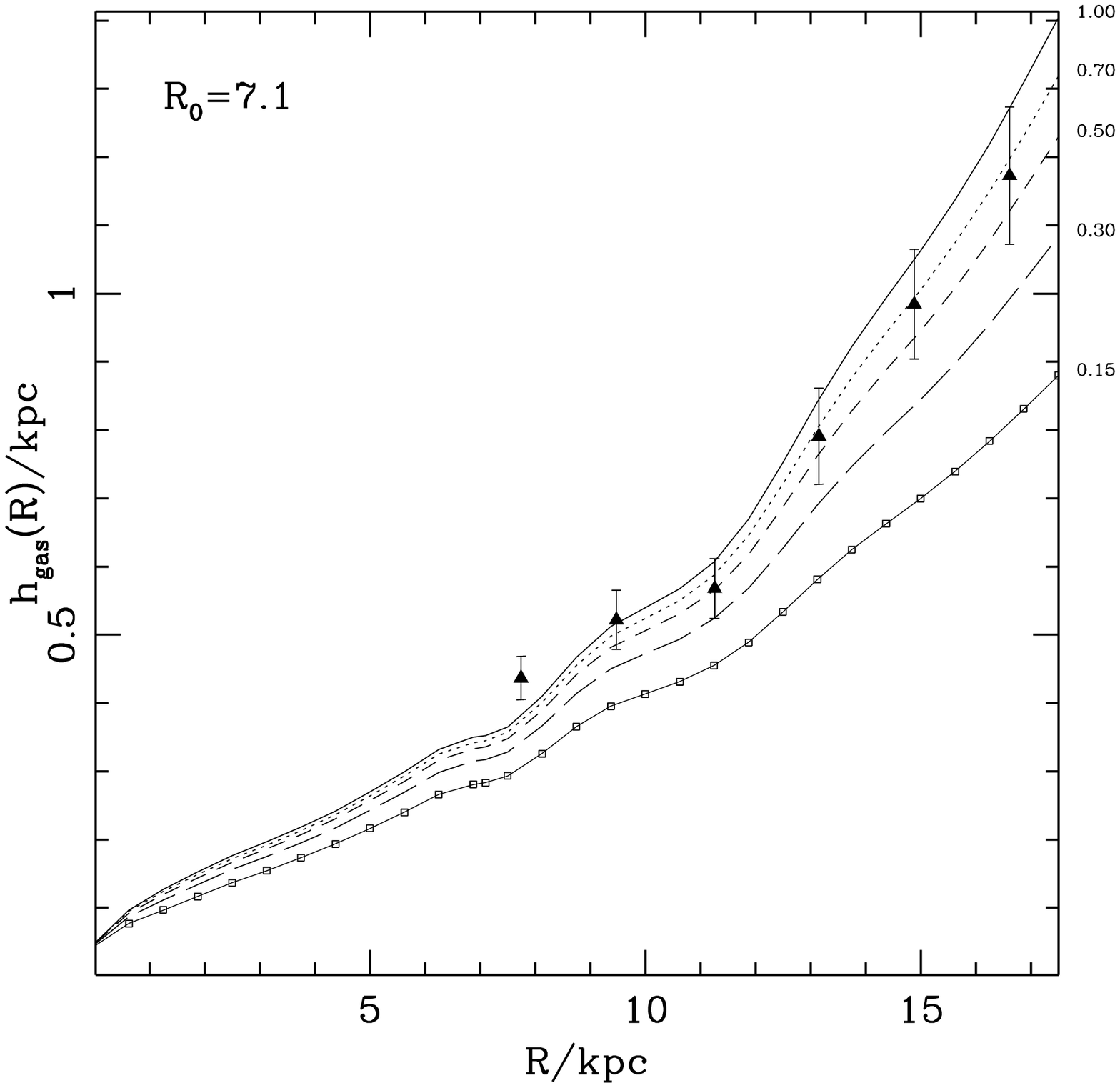}
\caption{Fits of hydrostatic equilibrium gas layer models to the
observed FWHM of the HI layer in the Milky Way for two different
possible values for $R_0$.  The data come from combining the analyses
of Diplas \& Savage (1991), Wouterloot et al.\ (1992) and Merrifield
(1992).  The fits were calculated using a wide range of plausible mass
models for the gravitational potential; the individual fits are
annotated by the value of $q$, the halo flattening of the model.}
\end{figure}

The procedure that we have adopted for estimating $q$ for the Milky
Way is therefore as follows:
\begin{enumerate}
\item Pick values for $R_0$ and $\Theta_0$ that do not violate the known
constraints on these quantities.
\item Calculate $h_{\rm gas}(R)$ and $\Theta(R)$ for these Galactic
constants using Merrifield's (1992) method at radii outside $R_0$, and
data from the tangent points (Malhotra 1995) for the inner Galaxy.
One slight caveat here is that the gas layer is observed using its
$21\,{\rm cm}$ emission, which is measured with radio telescopes whose
spatial resolutions are frequently comparable to the angular thickness
of the gas layer; it is therefore necessary to apply a correction for
this finite beam size when calculating the physical dimension of the
gas layer.
\item Calculate a mass model consistent with the observed rotation
curve.  This model is constructed by breaking the Galaxy up into the
usual components of a disk, a bulge, a gas layer, and a halo that is
assumed to take the form of a non-singular isothermal spheroid of
flattening $q$.
\item Use the equation of hydrostatic equilibrium to calculate how one
would expect the gas layer thickness to vary with radius in the
Galaxy.  In this analysis, it is important to include the self-gravity
of the gas layer, which significantly reduces its equilibrium
thickness.  The internal velocity dispersion of the gas is also an
important parameter in the calculation; here we adopt $\sigma =
9.2\,{\rm km}\,{\rm s}^{-1}$, as measured for the inner Galaxy (Malhotra
1995).  We have neglected the possibility that non-thermal pressures
due to cosmic rays and magnetic fields may provide extra support
against gravity, but it has been argued that these forces make
negligible contributions to the total pressure support of the gas
layer (Rupen 1991, Olling 1996).
\item Compare the model equilibrium gas layer thickness to the
observed $h_{\rm gas}(R)$.  If the data fit the model to within the
uncertainties, then the value for $q$ in this model is acceptable.
\item Repeat for all plausible values of $R_0$, $\Theta_0$ and $q$ to
see what range of values for $q$ is compatible with the observations.
\end{enumerate}
The practical application of this procedure to $21\,{\rm cm}$
observations of the Milky Way is described in more detail in Olling \&
Merrifield (1997).

The results of model fits to $h_{\rm gas}(R)$ are presented in Fig.~2.
Clearly, none of the models are consistent with the observations if we
adopt the IAU standard value of $R_0 = 8.5\,{\rm kpc}$ (Kerr \&
Lynden-Bell 1987).  However, by reducing $R_0$ toward the lower end of
the range allowed by existing measurements, it is possible to
reproduce the observed flaring of the gas layer quite accurately, even
down to the inflection at $\sim 12\,{\rm kpc}$.  The best-fit values
for $q$ decrease as $R_0$ is decreased, but for even the smallest
acceptable values of $R_0$ the inferred halo shape is quite round.
From the complete ensemble of models, we find that acceptable fits to
$h_{\rm gas}(R)$ can be obtained if $R_0 = 7 \pm 1\,{\rm kpc}$, and
that the corresponding uncertainty in the halo flattening is $q = 0.7
\pm 0.3$.

\section{The Shape of the Milky Way's Halo from Stellar Kinematics}

As mentioned in the Introduction, our location within the Milky Way
presents us with a further method for calculating the flattening of
the halo which is unique to this system.  Specifically, we can study
the kinematics of stars in the Galactic disk near the Sun.  The
motions of these stars perpendicular to the Galactic plane are
dictated by the local gravitational field, and so provide a measure of
the total mass located there.  Just as for the gas layer flaring
technique described above, this measure of the mass close to the plane
can be converted into a measure of halo flattening.  

Kuijken \& Gilmore (1991) used the spatial distribution and kinematics
of a sample of K dwarf stars in the solar neighborhood to carry out
such a measurement.  From these data, they found that the total mass
within $1.1\,{\rm kpc}$ of the Galactic plane can be robustly
estimated to be $\Sigma_{1.1} = 71 \pm 6\,M_\odot\,{\rm pc}^{-2}$.

Some fraction of this mass comes from the normal luminous stellar disk
of the Milky Way.  Clearly, in order to find the mass contribution
from the dark halo, we have to allow for this contribution to
$\Sigma_{1.1}$; Kuijken \& Gilmore (1989) calculated that the column
density contributed by the stellar component is $\Sigma_* = 35 \pm
5\,M_\odot\,{\rm pc}^{-2}$.

\begin{figure}
\plottwo{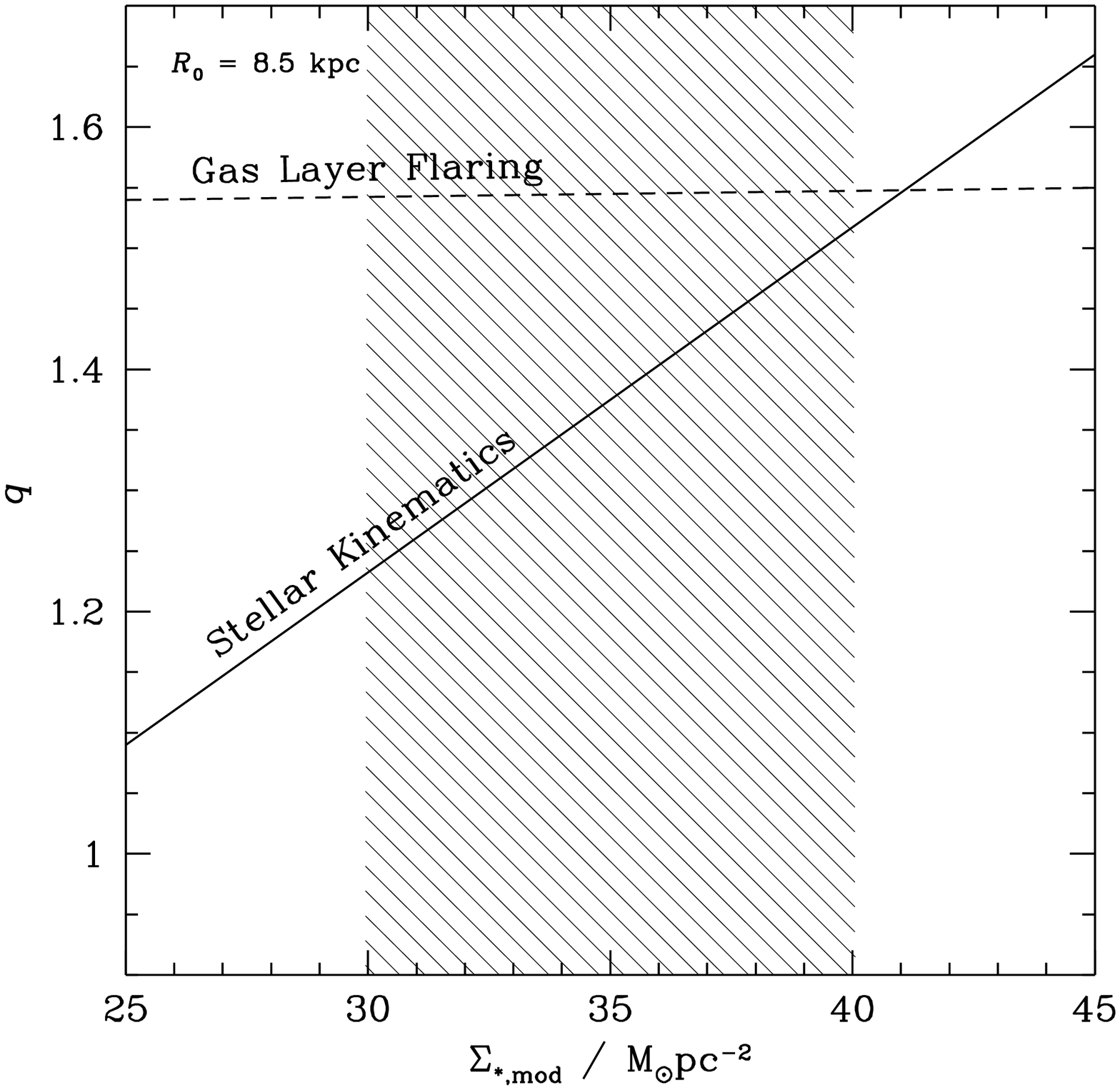}{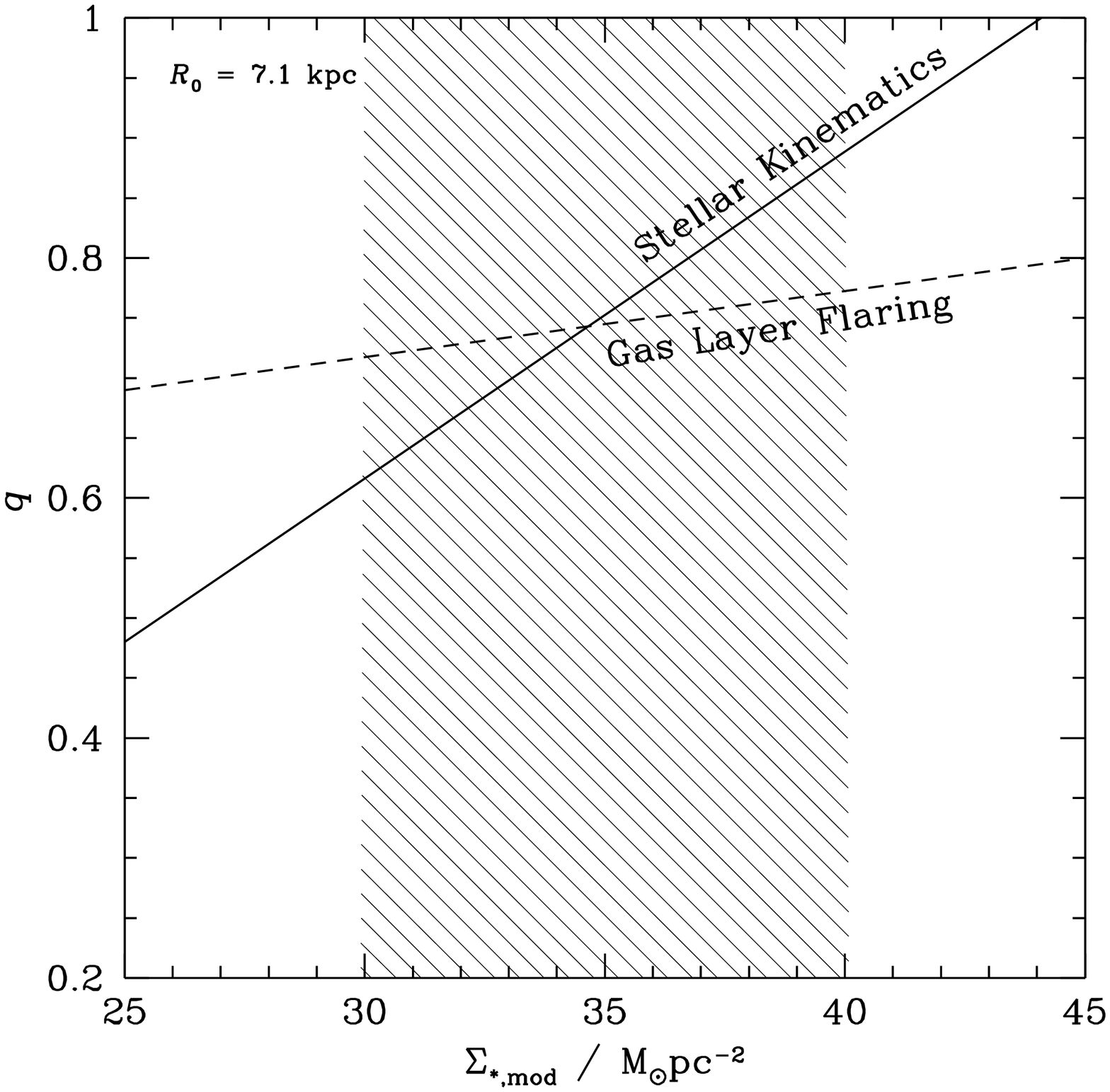}
\caption{The inferred flattening of the Galactic halo, $q$, as a
function of the adopted column of stellar disk mass in the solar
neighborhood, $\Sigma_{*,{\rm mod}}$, for two different adopted values
of $R_0$.  The shaded regions show the observationally-allowed range
for $\Sigma_*$.  The lines show the value of $q$ as derived from both
the stellar kinematic constraints and the gas layer flaring.}
\end{figure}

The procedure necessary to use these values of $\Sigma_{1.1}$ and
$\Sigma_*$ to estimate $q$ is essentially identical to that adopted
for the flaring analysis: we create mass models that reproduce the
observed rotation curve of the Milky Way, using values for $\Theta_0$
and $R_0$ consistent with the existing constraints.  We then calculate
the values of $\Sigma_{1.1}$ for these models, and reject those that
are inconsistent with the observed value.  The values of $q$ for the
acceptable models are plotted in Fig.~3.  These plots show $q$ as a
function of the stellar column density in the solar neighborhood as
calculated for each model, $\Sigma_{*,{\rm mod}}$.  These quantities
are strongly correlated: as $\Sigma_{*,{\rm mod}}$ is increased, it
accounts for more and more of $\Sigma_{1.1}$, so the contribution from
the dark halo must be correspondingly reduced.  Since the rotation
curve of the Milky Way is fixed (for given $R_0$ and $\Theta_0$), the
reduction in the contribution to $\Sigma_{1.1}$ from the dark halo
must be produced by making the halo rounder.  As fig.~3 illustrates,
there is a similar correlation between the values of $q$ derived from
gas flaring and $\Sigma_{*,{\rm mod}}$, but in this case the
dependence is much weaker, since the fit to the thickness of the gas
layer is made at larger radii, where the influence of the stellar disk
on the amount of mass close to the Galactic plane is greatly reduced.

Once again, the stellar kinematic analysis argues for a small value of
$R_0$.  As Fig.~3 shows, if we adopt $R_0 = 8.5\,{\rm kpc}$, the gas
flaring analysis and the stellar kinematic analysis only produce
consistent estimates for $q$ for models where $\Sigma_{*,{\rm mod}}$
lies outside the observed range of uncertainty for the local stellar
column density.  If, on the other hand, we set $R_0 = 7.1\,{\rm kpc}$,
we find that both methods produce consistent estimates of $q \approx
0.7$ for models in which $\Sigma_{*,{\rm mod}} \approx 35
\,M_\odot\,{\rm pc}^{-2}$, in good agreement with the observed value
of $\Sigma_*$.

\section{Discussion}

We have derived the flattening of the Milky Way's dark halo using two
entirely different techniques: analysis of the global flaring of its
HI gas layer with radius; and local stellar-kinematic studies of the
total mass in the solar neighborhood.  Both methods return consistent
results as long as we adopt values for the distance to the Galactic
center toward the lower end of the observationally-permitted range.
In fact, most of the error in the derived value of $q$ arises from the
current uncertainty in $R_0$.  If future, better determinations of
$R_0$ place the value closer to the IAU-sanctioned $8.5\,{\rm kpc}$,
the apparent consistency between the different estimators for $q$ will
have to be reconsidered.  With this proviso, we can combine the
results of the two analyses to obtain a best estimate for the
flattening of the Galactic halo of $q = 0.75 \pm 0.25$.

The consistency in the values of $q$ derived by the two different
methods goes some way toward justifying the various assumptions that
have been made in these analyses, such as the neglect of non-thermal
pressure support in calculating the degree of flaring in the gas
layer.  The value of $q \sim 0.7$ derived from the application of the
flaring technique to the Milky Way also shows that this method does
not necessarily always return small values for $q$.  The relatively
low values derived by this technique for other galaxies (see Fig.~1)
can therefore be put down to the very small number statistics.
These findings mean that we now have rather more confidence in the
application of the gas layer flaring technique to other galaxies.

The only set of results in Fig.~1 that is clearly inconsistent with a
mean value of $q \sim 0.5$ is that produced by the normal mode
analysis of warped gas disks.  The inconsistency with other techniques
means that this method must be viewed with some scepticism.  Indeed,
recent numerical work has shown that the normal modes invoked by
this method should be rapidly damped by dynamical friction (Dubinski
\& Kuijken 1995), suggesting that warps should not be described in
this manner.  If these points are neglected from Fig.~1, and the Milky
Way result from this paper is added, then the resulting data are
consistent with the simple CDM prediction shown.  It should be borne
in mind, though, that the data set is still embarrassingly small, and
many more measurements are required if one is to test this model
definitively.

However, even with the sparse data available, we can rule out some
other\-wise-cred\-ible possibilities.  As mentioned in the Introduction,
some scenarios invoke dark matter distributed in a disk-like
arrangement, for which one would expect to obtain consistently low
values of $q \sim 0.2$.  None of the available measurements of $q$
imply such flattened halos.  If any of these measurements are to be
believed, then cold molecular hydrogen (Pfenniger, Combes \& Martinet
1994) and decaying massive neutrinos (Sciama 1990) are no longer
plausible candidates for dynamically-significant amounts of dark
matter.

The analysis described in this paper also has implications for the
interpretation of microlensing experiments.  Microlensing events in
the Large Magellanic Cloud (LMC) are interpreted as arising from
lensing by compact massive dark objects in the halo of the Milky Way
(e.g.\ Alcock et al.\ 1997).  Clearly, the number of such objects
along the line of sight to the LMC depends not only on the radial
distribution of these objects in the Galaxy, but also on the degree of
flattening of flattened their distribution.  Thus, the measurement of
$q$ for the Milky Way has a bearing on the local determination of the
nature of dark matter, as well as the more global, cosmological
approach to this problem.

\acknowledgements MM is supported by a PPARC Advanced Fellowship 
(B/94/AF/1840).


\begin{references}
\reference Alcock, C., et al., 1997, ApJ, 486, 697
\reference Buote, D.A. \& Canizares, C.R., 1996, ApJ, 468, 184
\reference Diplas, A. \& Savage, D.S., 1991, ApJ, 377, 126
\reference Dubinski, J., 1994, ApJ, 431, 617
\reference Dubinski, J. \& Kuijken, K., 1995, ApJ, 442, 492
\reference Hofner, P. \& Sparke, L., 1994, ApJ, 428, 466
\reference Kerr, F.J. \& Lynden-Bell, D., 1986, MNRAS, 221, 1023
\reference Kuijken, K. \& Gilmore, G., 1989, MNRAS, 239, 605
\reference Kuijken, K. \& Gilmore, G., 1991, ApJ, 367, L9
\reference Malhotra, S., 1995, ApJ, 448, 138
\reference Merrifield, M.R., 1992, AJ, 103, 1552
\reference Navarro, J.F., Frenk, C.S. \& White, S.D.M., 1996, ApJ,
462, 563
\reference Olling, R.P., 1996, AJ, 112, 457
\reference Olling, R.P. \& Merrifield, M.R., 1997, in preparation
\reference Peebles, P.J.E., 1993, Principals of Physical Cosmology,
Princeton: Princeton University Press 
\reference Pfenniger, D., Combes, F. \& Martinet, L., 1994, A\&A, 285,
79 
\reference Reid, M.J., 1993, ARA\&A, 31, 345 
\reference Rupen, Mp>p, 1991, Ph.D. Thesis, Princeton University
\reference Sackett, P.D., Rix, H.-W., Jarvis, B.J. \& Freeman, K.C.,
1994, ApJ, 436, 629
\reference Sackett, P.D., 1997, ApJ, 483, 103
\reference Sciama, D., 1990, MNRAS, 244, 1 
\reference Steiman-Cameron, T.Y., Kormendy, J. \& Durisen, R.H., 1992,
AJ, 104, 1339 
\reference Wouterloot, J.G.A, Brand, J., Burton, W.B. \& Kwee, K.K., 1990,
A\&A. 230, 21
\end{references}
\end{document}